\documentclass{PoS}
\usepackage{amsmath}
\usepackage{cite}

\usepackage{tikz}
\newcommand{\hc}{\hbox {h.c.}}

\renewcommand{\Im}{\mbox{Im\thinspace}}

\title{CP Violation in the scalar sector}

\ShortTitle{CP Violation in the scalar sector}

\author{D. Emmanuel-Costa\\
        Centro de F\'\i sica Te\'orica de Part\'\i culas -- CFTP\\
Instituto Superior T\'ecnico -- IST, Universidade de Lisboa, Av. Rovisco Pais, \\
P-1049-001 Lisboa, Portugal\\
        E-mail: \email{david.costa@tecnico.ulisboa.pt}}
\author{O. M. Ogreid\\
        Bergen University College, Bergen, Norway\\
        E-mail: \email{omo@hib.no}}
\author{\speaker{P. Osland} \\
        University of Bergen, Norway\\
       E-mail: \email{per.osland@ift.uib.no}}

\author{M. N. Rebelo\\
        Centro de F\'\i sica Te\'orica de Part\'\i culas -- CFTP and Departamento de F\'\i sica\\
Instituto Superior T\'ecnico -- IST, Universidade de Lisboa, Av. Rovisco Pais, \\
P-1049-001 Lisboa, Portugal\\
        E-mail: \email{rebelo@tecnico.ulisboa.pt}}

\abstract{Models with an extended scalar sector may in principle provide new sources
of CP violation originating in the scalar potential.
One of the simplest ways to implement this idea is to have CP violation in a two-Higgs-doublet model. Here, it leads to CP violation in trilinear weak gauge boson couplings.
We discuss how these couplings are becoming constrained in the alignment limit.
In a model with three Higgs doublets, subject to an $S_3$ symmetry, several complex vacua are possible. Some of these, but not all, may lead to spontaneous CP violation.}

\FullConference{Proceedings of the Corfu Summer Institute 2015 "School and Workshops on Elementary Particle Physics and Gravity"\\
		1-27 September 2015\\
		Corfu, Greece}

\begin{document}

\section{Introduction}
In our attempts to identify physics beyond the standard model,
one may seek guidance from the fact that baryogenesis
\cite{Riotto:1999yt} requires additional CP violation.
Actually, it has long been known that the extension
of the standard model with an extra SU(2)$\times$U(1) scalar doublet
introduces additional sources of CP violation. In fact, models with two
Higgs doublets can violate CP either explicitly or spontaneously.
Spontaneous CP violation \cite{Lee:1973iz} 
has the attractive feature of putting on an equal footing CP and 
electroweak symmetry breaking. It should be stressed that 
spontaneous CP violation can only occur, provided 
the Lagrangian conserves CP. 
 
We shall here briefly review the constraints on CP violation in the
Two-Higgs-Doublet Model (2HDM) and then discuss a scalar potential with
three Higgs doublets. The general three-Higgs-doublet model has many
parameters, so we will restrict ourselves to the ten-parameter
$S_3$-symmetric potential. The additional, discrete, symmetry may also
provide a framework for dark matter.

The analysis presented here is important for model building. Symmetries
have the important feature of reducing the number of free parameters 
and at the same time leading to predictions that can in principle be verified
experimentally at the LHC.

\section{CP violation in the 2HDM}
One of the simplest models that allows for CP violation in the scalar sector is the Two-Higgs-Doublet Model (2HDM).
When the three neutral Higgs fields of this model mix, CP will be violated, either explicitly or spontaneously \cite{Lee:1973iz}. This mixing, which yields the three states $H_1$, $H_2$ and $H_3$, can be described by two additional mixing angles, replacing the familiar mixing angle $\alpha$ by a set of three, $(\alpha_1,\alpha_2,\alpha_3)$ \cite{Accomando:2006ga}. The regions in this parameter space where explicit and spontaneous CP violation can take place, have been discussed in Ref.~\cite{Grzadkowski:2013rza}. In fact, internal consistency and experimental constraints allow some amount of mixing, see, for example, Refs.~\cite{Basso:2012st,Basso:2013wna,Fontes:2015mea,Fontes:2015xva}.

From one point of view, CP violation is attractive since it may make baryogenesis possible, from another, it offers CP-violating observables \cite{Lavoura:1994fv,Botella:1994cs,Grzadkowski:2014ada} that one could try to measure or constrain experimentally. 
When neither of the three neutral Higgs bosons is an eigenstate of CP, then the neutral gauge boson will have trilinear couplings with all three pairs of neutral scalars. All these will also have trilinear coupings with the charged pair.

Conditions for a two-Higgs-doublet potential 
to conserve CP at the Lagrangian level expressed in terms of Higgs 
basis invariants and which are independent of the vacuum expectation 
values  were presented in Refs.~\cite{Branco:2005em}
and \cite{Gunion:2005ja}.
Within the bosonic sector of the 2HDM, i.e., without specifying the Yukawa couplings, CP-violating observables can all be expressed in terms of three invariants, $\Im J_1$, $\Im J_2$ and $\Im J_{30}$. These may in turn be expressed by the masses of the neutral sector, ($M_1,M_2,M_3$), as well as six couplings, in Ref.~\cite{Grzadkowski:2014ada} denoted $e_i$ and $q_i$, $i=1,2,3$.
Here, $e_i$ parametrizes the $ZZH_i$ coupling strength, as well as the $ZH_jH_k$ coupling (for $i\neq j\neq k\neq i$), whereas the $q_i$ parametrizes the $H^+H^-H_i$ coupling.
The quantity $\Im J_2$ actually induces interesting CP-violating effects in effective $ZZZ$ and $ZWW$ vertices
 \cite{Chang:1993vv,Chang:1994cs,Grzadkowski:2016lpv}.

While the data allow some amount of mixing,
recent data on the 125~GeV Higgs particle (assumed to be the lightest one, $H_1$) actually point to the decoupling limit \cite{Gunion:2002zf}, in which its couplings to the gauge bosons coincide with those of the Standard Model.
In particular, this implies
\begin{equation}
e_1=v, \quad e_2=0, \quad e_3=0,
\end{equation}
where $v=246~\text{GeV}$.
These values imply \cite{Grzadkowski:2014ada,Grzadkowski:2015zma}
\begin{align}
\Im J_1&=0, \nonumber \\
\Im J_2&=0, \nonumber \\
\Im J_{30}&=\frac{q_2q_3}{v^4}(M_3^2-M_2^2).
\end{align}
Recalling that $q_2$ and $q_3$ refer to the coupling strengths of a charged Higgs pair to the two heavier neutral ones, it is clear that it would be very challenging to try to measure the CP-violating quantity $\Im J_{30}$.

\section{CP violation in the  S$_3$-symmetric 3-Higgs-doublet model}
The $S_3$-symmetric three-Higgs-doublet potential, which is defined in terms of ten parameters, has a very rich structure. In the irreducible-representation framework (IRF), where the $SU(2)\times U(1)$ doublets $h_1$ and $h_2$ form an  $S_3$ doublet, whereas $h_S$ is an $S_3$ singlet,
it can be written as \cite{Das:2014fea}
\begin{align} \label{Eq:V-DasDey}
V&=\mu_0^2 h_S^\dagger h_S +\mu_1^2(h_1^\dagger h_1 + h_2^\dagger h_2) \nonumber \\
&+
\lambda_1(h_1^\dagger h_1 + h_2^\dagger h_2)^2 
+\lambda_2(h_1^\dagger h_2 - h_2^\dagger h_1)^2
+\lambda_3[(h_1^\dagger h_1 - h_2^\dagger h_2)^2+(h_1^\dagger h_2 + h_2^\dagger h_1)^2]
\nonumber \\
&+ \lambda_4[(h_S^\dagger h_1)(h_1^\dagger h_2+h_2^\dagger h_1)
+(h_S^\dagger h_2)(h_1^\dagger h_1-h_2^\dagger h_2)+\hc] 
+\lambda_5(h_S^\dagger h_S)(h_1^\dagger h_1 + h_2^\dagger h_2) \nonumber \\
&+\lambda_6[(h_S^\dagger h_1)(h_1^\dagger h_S)+(h_S^\dagger h_2)(h_2^\dagger h_S)] 
+\lambda_7[(h_S^\dagger h_1)(h_S^\dagger h_1) + (h_S^\dagger h_2)(h_S^\dagger h_2) +\hc]
\nonumber \\
&+\lambda_8(h_S^\dagger h_S)^2.
\end{align}
The same ten-parameter potential can also be represented in a complementary, reducible-repre\-sentation framework (RRF), where the three $SU(2)\times U(1)$ doublets $(\phi_1,\phi_2,\phi_3)$ are treated on an equal footing \cite{Derman:1978rx}. There is a linear mapping between these two potentials, as physical models they are thus equivalent until some other sector is specified, like for example Yukawa couplings.

A couple of features of this potential are worth stressing:
\begin{itemize}
\item
The potential is invariant under $h_1\to-h_1$, but {\it not} under $h_2\to-h_2$.
\item
For $\lambda_4=0$, the potential has an additional SO(2) symmetry.
In addition, the potential is then invariant under $h_2\to-h_2$, and under $h_1\leftrightarrow h_2$.
\end{itemize}

Recently, a complete catalogue of possible vacua has been given \cite{Emmanuel-Costa:2016vej}, with an emphasis on the complex ones and the corresponding constraints on the parameters of the potential. 
We list the complex vacua in Table~\ref{Table:complex}. The IRF specification, in terms of vacuum expectation values (vevs) $(w_1,w_2,w_S)$ is given for all of them, whereas the corresponding RRF specification in terms of the vevs $(\rho_1,\rho_2,\rho_3)$ is only given for the simpler cases.
The minima are determined by solving minimization conditions for three moduli and two relative phases, a total of five conditions. For most of the vacua, these five conditions are not independent. Indeed, the roman numeral making up the middle element (I, III, etc) of the vacuum name given in Table~\ref{Table:complex} refers to the number of independent conditions.

\begin{table}[htb]
\caption{Complex vacua (after Ref.~\cite{Emmanuel-Costa:2016vej}). Symbols with a ``hat'' (like $\hat w_S$) are real and positive. The vacua labelled with a checkmark (\checkmark) violate CP spontaneously, whereas those labelled with an asterisk ($^\ast$) are in fact real, due to the constraints that have to be imposed.}
\label{Table:complex}
\begin{center}
\begin{tabular}{|c|c|c|c|}
\hline\hline
Name& IRF (Irreducible Rep.)& RRF  (Reducible Rep.) & SCPV \\
\hline
& $w_1,w_2,w_S$ & $\rho_1,\rho_2,\rho_3$ &  \\
\hline
\hline
C-I-a & $\hat w_1,\pm i\hat w_1,0$ & 
$x, xe^{\pm\frac{2\pi i}{3}}, xe^{\mp\frac{2\pi i}{3}}$ & \\
\hline
\hline
C-III-a & $0,\hat w_2e^{i\sigma_2},\hat w_S$ & $y, y, xe^{i\tau}$  & \checkmark \\
\hline
C-III-b & $\pm i\hat w_1,0,\hat w_S$ & $x+iy,x-iy,x$  & \\
\hline
C-III-c & $\hat w_1 e^{i\sigma_1},\hat w_2e^{i\sigma_2},0$ 
&   & \checkmark \\
\hline
C-III-d & $\pm i \hat w_1,\hat w_2,\hat{w}_S$ & $xe^{ i\tau},xe^{- i\tau},y$ & \\
\hline
C-III-e & $\pm i \hat w_1,-\hat w_2,\hat{w}_S$ & $xe^{ i\tau},xe^{- i\tau},y$ & \\
\hline
C-III-f & $\pm i\hat w_1 ,i\hat w_2,\hat{w}_S$ 
&  & \\
\hline
C-III-g & $\pm i\hat w_1,-i\hat w_2,\hat{w}_S$ 
&  & \\
\hline
C-III-h & $\sqrt{3}\hat w_2 e^{i\sigma_2},\pm\hat w_2 e^{i\sigma_2},\hat{w}_S$ 
& $xe^{i\tau} , y , y$, \quad
$y, xe^{i\tau},y$ & \checkmark \\
\hline
C-III-i & $\sqrt{\frac{3(1+\tan^2\sigma_1)}{1+9\tan^2\sigma_1}}\hat w_2e^{i\sigma_1},$ 
& $x, ye^{i\tau},ye^{-i\tau}$ & \\
& $\pm\hat w_2e^{-i\arctan(3\tan\sigma_1)},\hat w_S$ 
& $ye^{i\tau}, x, ye^{-i\tau}$ &  \\
\hline
\hline
C-IV-a$^\ast$ & $\hat w_1e^{i\sigma_1},0,\hat w_S$ &  & \\
\hline
C-IV-b & $\hat w_1,\pm i\hat w_2,\hat w_S$ 
& & \\
\hline
C-IV-c & $\sqrt{1+2\cos^2\sigma_2}\hat w_2,$ & & \\
& $\hat w_2e^{i\sigma_2},\hat w_S$ 
&  & \checkmark \\
\hline
C-IV-d$^\ast$ & $\hat w_1e^{i\sigma_1},\pm\hat w_2e^{i\sigma_1},\hat w_S$ & & \\
\hline
C-IV-e & $\sqrt{-\frac{\sin 2\sigma_2}{\sin 2\sigma_1}}\hat w_2e^{i\sigma_1},$ & & \\
& $\hat w_2e^{i\sigma_2},\hat w_S$ & & \checkmark \\
\hline
C-IV-f & $\sqrt{2+\frac{\cos \left(\sigma _1-2 \sigma _2\right)}{\cos\sigma_1}}\hat w_2e^{i\sigma_1},$ & & \\
& $\hat w_2e^{i\sigma_2},\hat w_S$ & & \checkmark \\
\hline
\hline
C-V$^\ast$ & $\hat w_1e^{i\sigma_1},\hat w_2e^{i\sigma_2},\hat w_S$ & $xe^{i\tau_1},ye^{i\tau_2},z$ & \\
\hline
\end{tabular}
\end{center}
\end{table}

Many of the complex vacua support spontaneous CP violation. However, the model can also lead to complex vacua that do not violate CP. This is due to the symmetry of the potential, as will be shown by a few examples.

\subsection{Example: C-I-a}
This is a well-known case \cite{Branco:1983tn}, best discussed in the reducible-representation framework.
Under complex conjugation (c.\,c.), the vacuum undergoes the following transformation:
\begin{equation}
(\rho_1,\rho_2,\rho_3)=
(x, xe^{\pm 2\pi i/3}, xe^{\mp2\pi i/3})
{\buildrel{\text{c.c.}}\over{\longrightarrow}} (x, xe^{\mp2\pi i/3}, xe^{\pm 2\pi i/3}).
\end{equation}
The vevs of $\phi_2$ and $\phi_3$ have been complex conjugated. However, the potential is symmetric under the interchange of $\phi_2$ and $\phi_3$, and since the moduli are the same, it remains invariant under complex conjugation.

\subsection{Example: C-III-a}
In the RRF, this has the form
\begin{equation}
(\rho_1,\rho_2,\rho_3)=(y,y,xe^{i\tau})
{\buildrel{\text{c.c.}}\over{\longrightarrow}}
(y,y,xe^{-i\tau}),
\end{equation}
whereas in the IRF it has the form
\begin{equation}
(w_1,w_2,w_S)=(0,\hat w_2e^{i\sigma_2},\hat w_S)
{\buildrel{\text{c.c.}}\over{\longrightarrow}}
(0,\hat w_2e^{-i\sigma_2},\hat w_S),
\end{equation}
In this case, no symmetry operation can ``undo'' a complex conjugation, and CP is spontaneously violated.

\subsection{Example: C-III-b}
In the IRF this has the form
\begin{equation}
(w_1,w_2,w_S)=(\pm i\hat w_1,0,\hat w_S)
{\buildrel{\text{c.c.}}\over{\longrightarrow}}
(\mp i\hat w_1,0,\hat w_S).
\end{equation}
At first sight, it looks like this would lead to CP violation. However, we recall that the potential is invariant under the interchange $h_1\to -h_1$, so there is no CP violation. The same conclusion is also easily reached in the RRF, and applies also to C-III-d and C-III-e.

\subsection{Example: C-III-c}
In the IRF this has the form
\begin{equation}
(w_1,w_2,w_S)=(\hat w_1 e^{i\sigma_1},\hat w_2 e^{i\sigma_2},0)
{\buildrel{\text{c.c.}}\over{\longrightarrow}}
(\hat w_1 e^{-i\sigma_1},\hat w_2 e^{-i\sigma_2},0).
\end{equation}
Since $w_S=0$, a rephasing allows for the removal of one phase:
\begin{equation}
(w_1,w_2,w_S)=(\hat w_1 e^{i\sigma},\hat w_2 e^{-i\sigma},0)
{\buildrel{\text{c.c.}}\over{\longrightarrow}}
(\hat w_1 e^{-i\sigma},\hat w_2 e^{i\sigma},0).
\end{equation}
This vacuum requires $\lambda_4=0$, in which case the potential is symmetric under the interchange $h_1\leftrightarrow h_2$. However, in the general case, for $\hat w_1\neq \hat w_2$, this vacuum leads to CP violation.

We proposed adding a soft term to avoid massless neutral scalars in
this case. In section 9 of our paper \cite{Emmanuel-Costa:2016vej} we did not introduce the
most general soft breaking terms and we incorrectly state that
with our choice there is still CP violation.

\subsection{Example: The Pakvasa--Sugawara vacuum}
The following complex vacuum was identified by Pakvasa and Sugawara many years ago \cite{Pakvasa:1977in}:
\begin{equation}
(w_1,w_2,w_S)=(\hat w e^{i\sigma},\hat w e^{-i\sigma},\hat w_S).
\end{equation}
Superficially, this looks like it might lead to CP violation. However, for consistency, it requires $\lambda_4=0$, in which case the potential is symmetric under the interchange $h_1\leftrightarrow h_2$. Hence, there is no CP violation. (This vacuum is contained in C-III-f, C-III-g and C-IV-e, depending on which additional conditions are imposed, in addition to $\lambda_4=0$.)
\subsection{Example: The Ivanov--Nishi vacuum}
The following complex vacuum was identified by Ivanov and Nishi \cite{Ivanov:2014doa}:
\begin{equation}
(w_1,w_2,w_S)=(\hat w e^{i\sigma},\hat w e^{i\sigma},\hat w_S)
\end{equation}
Also this one requires $\lambda_4=0$.
Additional conditions on this solution imply that it does not lead to CP violation.
In fact, it is a special case of C-III-f, C-III-g or C-IV-d, depending on which additional conditions are imposed, together
with $\lambda_4 = 0$.

\begin{center}
----------------------
\end{center}
Note that the vacuum
\begin{equation} \label{Eq:example_RRF}
(\rho_1,\rho_2,\rho_3)=x(e^{i\tau},e^{i\tau},1),
\end{equation}
which is a special case of C-III-a (with $y=x$), obtained after a complex conjugation and an overall phase rotation by $e^{i\tau}$,
violates CP,
whereas
\begin{equation} \label{Eq:example_IRF}
(w_1,w_2,w_S)=\hat w(e^{i\sigma},e^{i\sigma},1)
\end{equation}
does not. It is just a special case of the Ivanov--Nishi vacuum.
While these two vacua, Eqs.~(\ref{Eq:example_RRF}) and (\ref{Eq:example_IRF}) have the same form, the important difference, which leads to opposite conclusions about CP violation, is the fact that they refer to different frameworks. The two frameworks represent different symmetries among the three fields.

A detailed discussion of all the vacua of Table~\ref{Table:complex} is given in Ref.~\cite{Emmanuel-Costa:2016vej}.
\section{Concluding remarks}

We have discussed two important features of multi-Higgs models.
These are: the fact that such models may provide new sources of
CP violation as well as good dark matter candidates. Sources
of CP violation beyond the SM are required to explain the observed
baryon asymmetry of the Universe and their effects may be observed
soon at the LHC or in future colliders. New sources of CP
violation may manifest themselves both in the scalar, the gauge  and in the
flavour sectors. The recently discovered Higgs boson at the LHC
has been under intense experimental study and it looks as if
it may closely behave as a standard-like Higgs boson. However,
on one hand, there is still room for deviations from SM couplings
for the discovered boson and on the other hand, these models predict
additional scalars, which may soon be discovered. At present, there is
a  hint for a new 750 GeV boson both from ATLAS and CMS \cite{750GeV}.
The nature of dark matter is another puzzle constituting one of the
most important open questions in our field.

Extensions of the scalar sector allow for a large number of new
parameters. Symmetries play the r\^ ole of reducing this number
and at the same time of establishing connections among different
phenomena.  We have seen that the $S_3$ symmetric potential has a
very rich structure. Some of these vacua require $\lambda_4=0$ for
consistency reasons. In this case the potential acquires an
additional SO(2) symmetry as mentioned in section 3. Spontaneous
breaking of this continuous symmetry would then lead to scalar
massless states which are  experimentally ruled out.
One possible way out is to include soft terms in the Higgs potential
breaking this symmetry. Soft breaking terms may also have
interesting implications for spontaneous CP violation as pointed out long
ago \cite{Branco:1985aq}. Some of the vacua listed in Table~\ref{Table:complex}
have vanishing vevs  for some fields. When endowed with a stabilizing symmetry,
like $Z_2$, for example, those fields might represent dark matter.
\bigskip

\noindent
{\bf Acknowledgements: }
We thank the local organizers of Corfu2015 for the very fruitful
scientific meeting and the warm hospitality.
The work of DE-C and MNR was partially supported by Funda\c{c}\~ao
para a Ci\^encia e a Tecnologia (FCT, Portugal)
through the projects CERN/FIS-NUC/0010/2015, and
CFTP-FCT Unit 777 (UID/FIS/00777/2013) which are partially
funded through POCTI (FEDER), COMPETE, QREN and EU.
The work of PO was supported in part by the Research Council of Norway.

\end{document}